\documentclass[aps,prd,12pt,notitlepage,nofootinbib,tightenlines]{revtex4-1}
\usepackage{amsmath}
\usepackage{bm}
\usepackage{times}
\usepackage{braket}
\usepackage{color}
\usepackage{epsfig}
\usepackage{slashed}
\usepackage{hyperref}
\newcommand{\beq}{\begin{eqnarray}}
\newcommand{\eeq}{\end{eqnarray}}
\newcommand{\non}{\nonumber\\ }

\newcommand{\etar}{\eta^\prime }
\newcommand{\etap}{\eta^{(\prime)} }

\newcommand{\cala}{ {\cal A} }

\newcommand{\bszb}{ \bar{B}_s^0}

\newcommand{\mbs}{m_{B_s} }

\newcommand{\jpsi}{J/\Psi}

\newcommand{\psl}{ P \hspace{-2.4truemm}/ }
\newcommand{\nsl}{ n \hspace{-2.2truemm}/ }
\newcommand{\vsl}{ v \hspace{-2.2truemm}/ }

\newcommand{\calh}{ {\cal H} }
\newcommand{\ov}{ \overline  }

\def \cpc{ Chin. Phys. C  }

\def \csb{ Chin. Sci. Bull. }
\def \ctp{ Commun. Theor. Phys.  }
\def \epjc{ Eur. Phys. J. C }
\def \jpg{  J. Phys. G }
\def \npb{  Nucl. Phys. B }
\def \plb{  Phys. Lett. B }
\def \ppnp{ Prog.Part. $\&$ Nucl. Phys. }
\def \prd{  Phys. Rev. D }
\def \prl{  Phys. Rev. Lett.  }
\def \zpc{  Z. Phys. C }
\def \jhep{ JHEP }

\definecolor{Red}{rgb}{1.,0.,0.}

\definecolor{Blue}{rgb}{0.,0.,1.}

\definecolor{nicered}{rgb}{0.7,0.1,0.1}
\definecolor{nicegreen}{rgb}{0.1,0.5,0.1}

\hypersetup{colorlinks,citecolor=nicegreen,linkcolor=nicered}
\begin{document}
%%%%%%%%%%%%%%%%%%%%%%%%%%%%%%%%%%%%%%%%%%%%%%%%
\title{$\bar{B}^0_s \to (\pi^0 \etap, \etap\etap)$ decays and the effects of next-to-leading order
contributions in the perturbative QCD approach}
\author{Zhen-Jun Xiao$^{1,2}$}\email{xiaozhenjun@njnu.edu.cn}
\author{Ya Li$^1$, Dong-Ting Lin$^1$, Ying-Ying Fan$^3$, and Ai-Jun Ma$^1$} %%
\affiliation{1. Department of Physics and Institute of Theoretical Physics,
Nanjing Normal University, Nanjing, Jiangsu 210023, P.R. China}
\affiliation{2. Jiangsu Key Laboratory for Numerical Simulation of Large Scale
Complex Systems, Nanjing Normal University, Nanjing 210023, P.R. China}
\affiliation{3. College of Physics and Electronic Engineering,
Xinyang Normal University, Xinyang, Henan 464000, P.R. China}
\date{\today}
%%--------------------------------------------------------------------
\begin{abstract}
In this paper, we calculate the branching ratios and CP violating asymmetries
of the five $\bar{B}^0_s \to (\pi^0\etap,\etap\etap)$ decays, by employing the perturbative QCD (pQCD) factorization
approach and with the inclusion of all
currently known next-to-leading order (NLO) contributions.
We find  that (a) the NLO contributions can provide
about $100\%$  enhancements to the LO pQCD predictions for the decay rates
of $\bar{B}_s^0 \to \eta\etar$ and $\etar\etar$ decays, but result in  small changes
to $Br(\bar{B}_s \to \pi^0 \etap)$ and $Br(\bar{B}_s \to \eta\eta)$;
(b) the newly known NLO twist-2 and twist-3 contributions to the relevant
form factors can provide about $10\%$ enhancements to the decay rates of the considered decays;
(c) for $\bar{B}_s \to \pi^0 \etap$ decays, their direct CP-violating
asymmetries $\cala_f^{dir}$ could be enhanced significantly by the inclusion of the NLO contributions;
and (d) the pQCD predictions for $Br(\bar{B}_s \to \eta \etap)$ and $Br(\bar{B}_s \to \etar\etar)$
can be as large as $4\times 10^{-5}$, which may be measurable at LHCb or the forthcoming
super-B experiments.
\end{abstract}
%%---------------------------------------------------------------------
 \vspace{1cm}

\maketitle
%%{\bf \rm Key Words:}{$B/B_s$ meson semileptonic decays; The pQCD factorization approach;
%% Form factors; Branching ratios; LHCb experiments}

%%------------------------------------------------------------------------

\section{Introduction}

As is well-known, the studies for the mixing and decays of $B_s$ meson play an
important role in testing the standard model (SM) and in searching for the
new physics beyond the SM \cite{bphys1,csbbs}.
Some $B_s$ meson decays, such as the leptonic decay $B_s^0 \to \mu^+\mu^-$ and
the hadronic decays $B_s^0 \to (\jpsi \phi, \phi\phi, K \pi, KK, etc)$, have been measured
recently by the LHCb, ATLAS and CMS collaborations \cite{lhcb1,lhcb2,lhcgrs}.

In a very recent paper \cite{xiao145}, we studied the $\bszb \to (K\pi,KK)$ decays by employing the pQCD
factorization approach with the inclusion of the NLO contributions
\cite{nlo05,xiao08a,xiao08b,fan2013,prd85-074004,cheng14a,xiao2014} and
found that the NLO contributions can interfere with the leading order (LO)
part constructively or destructively for
different decay modes, and can improve the agreement between the SM
predictions and the measured values for the considered decay modes \cite{xiao145}.
The charmless hadronic two-body decays of $B_s$ meson, in fact,
have been studied intensively by many authors by using rather different theoretical methods:
such as the generalized factorization \cite{chenbs99,xiaobs01},
the QCD factorization (QCDF) approach \cite{npb675,sun2003,chengbs09}
and the pQCD  factorization  approach at the LO or partial NLO level 
\cite{bspipi,pieta,ctp08,ali07,xiao08a}.
In Refs.\cite{xiao08b,fan2013,nlo05,xiao2014}, the authors proved that the NLO
contributions can play a key role in understanding the very large
$Br(B \to K \eta^\prime)$ \cite{xiao08b,fan2013}, the so-called ``$K\pi$-puzzle"
\cite{nlo05,xiao2014}, and the newly observed branching ratios and
CP violating asymmetries of $B_s\to K^+\pi^-$ and $B_s\to K^+ K^-$ decays \cite{lhcb1,lhcb2,xiao145}.

In this paper, we will calculate the branching ratios and CP violating asymmetries of the
five $B^0_s \to (\pi^0, \etap)\etap$ decays  by employing the pQCD approach.
We focus on the studies for the effects of various NLO contributions to the five $\bszb \to (\pi^0
\etap, \eta\eta,\eta\etar,\etar \etar)$ decays, specifically those NLO twist-2 and twist-3
contributions to the form factors of $B_s^0 \to \pi, \etap$ transitions \cite{prd85-074004,cheng14a}.

\section{ Decay amplitudes at LO and NLO level}\label{sec:lo-nlo}

As usual, we treat the $B_s$ meson as a heavy-light system and considered it at
rest for simplicity. Using the light-cone coordinates, we define the emitted meson $M_2$ moving along the direction
of $n=(1,0,{\bf 0}_{\rm T})$ and another meson $M_3$ the direction of $v=(0,1,{\bf 0}_{\rm T})$, and we also use $x_i$
to denote the momentum fraction of anti-quark in each meson:
\beq
P_{B_s} &=& \frac{\mbs}{\sqrt{2}} (1,1,{\bf 0}_{\rm T}), \quad
P_2 = \frac{M_{B_s}}{\sqrt{2}}(1,0,{\bf 0}_{\rm T}), \quad
P_3 = \frac{M_{B_s}}{\sqrt{2}} (0,1,{\bf 0}_{\rm T}),\non
k_1 &=& (x_1 P_1^+,0,{\bf k}_{\rm 1T}), \quad
k_2 = (x_2 P_2^+,0,{\bf k}_{\rm 2T}), \quad
k_3 = (0, x_3 P_3^-,{\bf k}_{\rm 3T}).
\eeq
After making the integration over $k_1^-$, $k_2^-$, and $k_3^+$ we find the conceptual decay amplitude
\beq
\cala &\sim &\int\!\! d x_1 d x_2 d x_3 b_1 d b_1 b_2 d b_2 b_3 d b_3 \non && \cdot \mathrm{Tr}
\left [ C(t) \Phi_{B_s}(x_1,b_1) \Phi_{M_2}(x_2,b_2) \Phi_{M_3}(x_3, b_3) H(x_i,
b_i, t) S_t(x_i)\, e^{-S(t)} \right ], \quad \label{eq:a2}
\eeq
where $b_i$ is the conjugate space coordinate of $k_{\rm iT}$,  $C(t)$ are the Wilson
coefficients evaluated at the scale $t$, and $\Phi_{B_s}$ and $\Phi_{M_i}$ are wave functions
of the $B_s$ meson and the final state mesons. The Sudakov factor $e^{-S(t)}$ and $S_t(x_i)$ together
suppress the soft dynamics effectively \cite{li2003}.

For the considered $B_s$ decays with a quark level transition $b \to q'$ with $q'=(d,s)$,
the weak effective Hamiltonian $H_{eff}$ can be written as \cite{buras96}
\beq
\label{eq:heff}
\calh_{eff} = \frac{G_{F}} {\sqrt{2}}\, \sum_{q=u,c}V_{qb} V_{qq'}^*\left\{  \left [ C_1(\mu) O_1^q(\mu)
+ C_2(\mu) O_2^q(\mu) \right ] + \sum_{i=3}^{10} C_{i}(\mu) \;O_i(\mu) \right\} \; .
\eeq
where $G_{F}=1.166 39\times 10^{-5}$ GeV$^{-2}$ is the Fermi constant, and
$V_{ij}$ is the Cabbibo-Kobayashi-Maskawa (CKM) matrix element, $C_i(\mu)$ are the Wilson coefficients and $O_i(\mu)$
are the four-fermion operators.

For $B_s^0$ meson, we consider only the contribution of Lorentz structure
\beq
\Phi_{B_s}= \frac{1}{\sqrt{2N_c}} (\psl_{B_s} +m_{B_s}) \gamma_5 \phi_{B_s} ({\bf k_1}),
\label{eq:bsmeson}
\eeq
with the distribution amplitude widely used in literature\cite{bspipi,pieta,ali07,xiao08a,xiao145}
\beq
\phi_{B_s}(x,b)&=& N_{B_s} x^2(1-x)^2 \exp \left  [ -\frac{M_{B_s}^2\ x^2}{2 \omega_{B_s}^2} -\frac{1}{2} (\omega_{B_s} b)^2\right],
\label{phib}
\eeq
where the parameter $\omega_{B_s}$ is a free parameter and we take $\omega_{B_s} =0.50 \pm 0.05$ GeV for $B_s$
meson. For fixed $\omega_{B_s}$ and $f_{B_s}$, the normalization factor $N_{B_s}$
can be determined through the
normalization condition: $\int\frac{d^4 k_1}{(2\pi)^4}\phi_{B_s}({\bf k_1}) =f_{B_s}/(2\sqrt{6})$.

For the light $\pi, K, \eta_q$ and $\eta_s$, their wave functions are similar in form and can be defined as
in Refs.~\cite{pball98,pball06,csbwf}
\beq
\Phi(P,x,\zeta)\equiv \frac{1}{\sqrt{2N_C}}\gamma_5 \left [ \psl \phi^{A}(x)+m_0
\phi^{P}(x)+ \zeta m_0 (\nsl \vsl -1)\phi_{P}^{T}(x)\right ],
\label{eq:phi-x1}
\eeq
where $P$ and $m_0$ are the momentum and the chiral mass of the light mesons.
When the momentum fraction of the quark (anti-quark) of the meson is set to be
$x$, the parameter $\zeta$ should be chosen as $+1$ ($-1$). The
expressions of the relevant twist-2 ($\phi^{A}(x)$)
and twist-3 ($\phi^{P,T}(x)$) distribution amplitudes of the mesons
$M=(\pi, K, \eta_q, \eta_s)$ and the relevant
chiral masses can be found easily in Refs.\cite{fan2013,xiao145}.
The relevant Gegenbauer moments $a_i$ have been chosen as in Ref.~\cite{ali07}:
\beq
a^{\pi, \eta_q, \eta_s}_1=0,\quad a_2^{\pi, \eta_q, \eta_s}=0.44\pm 0.22.
\eeq
The values of other parameters are $\eta_3=0.015$ and $\omega=-3.0$.

For the $\eta$-$\eta^\prime$ system, we use the traditional quark-flavor mixing
scheme: the physical states $\eta$ and $\etar$ are related to the flavor states
$\eta_q= (u\bar u +d\bar d)/\sqrt{2}$  and $\eta_s=s\bar{s}$ through a single mixing
angle $\phi$,
\beq
\eta =  \cos{\phi}\; \eta_q -\sin{\phi}\; \eta_s, \qquad \etar =  \sin{\phi}\; \eta_q +\cos{\phi}\;
\eta_s. \label{eq:e-ep}
\eeq
The relation between the decay constants $(f_q,f_s)$ and $(f_\eta^q,
f_\eta^s,f_{\etar}^q,f_{\etar}^s)$, as well as the chiral enhancement
$m_0^q$ and $m_0^s$, have been defined for example in
Ref.~\cite{fan2013}. The parameters $f_q, f_s$ and $\phi$ have been
extracted from the data \cite{fks98}:
 \beq
f_q=(1.07\pm 0.02)f_{\pi},\quad f_s=(1.34\pm 0.06)f_{\pi},\quad \phi=39.3^\circ\pm 1.0^\circ,
 \eeq
with $f_\pi=130$~MeV.

%%%%%%%%%%%%%%%%%%%%%%%%%%%%%%%%%%%%%%%%%%%%%%%%%%%%%%%%%%%%%%%%%%%%%%%%%%%%%%%%%%%%%%%%%
%%%%%%%%%%%%%%%%%%%%%%%%%%%%%%%%%%%%%%%%%%%%%%%%%%%%%%%%%%%%%%%%%%%%%%%%%%%%%%%%%%%%%%%%%

\subsection{ LO amplitudes}\label{sec:lo-aml}

\begin{figure}[tb]
\vspace{-5cm}
\centerline{\epsfxsize=18cm \epsffile{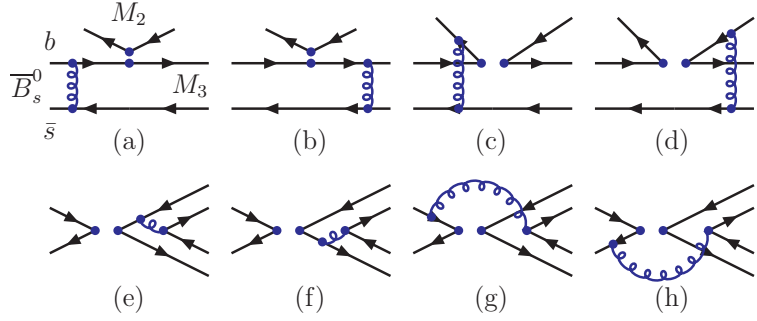}}
\vspace{-16cm}
\caption{ Feynman diagrams which may contribute at leading order to $B_s^0 \to (\pi^0,\etap)\etap$ decays.}
 \label{fig:fig1}
\end{figure}

The five $B_s^0 \to \pi^0 \etap, \eta\eta,\etar\etar,\eta\etar$ decays considered in this paper
have been studied previously in Ref.~\cite{pieta,ali07} by employing the pQCD
factorization approach at the leading order.
The decay amplitudes as presented in Ref.~\cite{pieta,ali07} are confirmed by our
recalculation. We here focus on the examination for the possible effects of all currently known
NLO contributions to these five decay modes in the pQCD factorization approach.
The relevant Feynman diagrams  which may contribute to the considered $B_s^0$
decays at the leading order are  illustrated in Fig.~\ref{fig:fig1}. We firstly show the LO decay amplitudes.

For $\bar {B}_s^0\to \pi^0 \etap$ decays, the LO decay amplitudes are
 \beq
  {\cal A}({\bar B}_s^0\to \pi^0\eta) &=&{\cal A}({\bar B}_s^0\to\pi^0\eta_q) \cos\phi
-{\cal A}({\bar B}_s^0\to\pi^0\eta_s) \sin\phi,\label{eq:pi01}\\
{\cal A}({\bar B}_s^0\to\pi^0\etar)&=& {\cal A}({\bar B}_s^0\to\pi^0\eta_q) \sin\phi
 +{\cal A}({\bar B}_s^0\to\pi^0\eta_s) \cos\phi,\label{eq:pi02}
 \eeq
with
 \beq
{\cal A}({\bar B}_s^0\to\pi^0\eta_q)&=&     \xi_u \left ( f_{B_s}F_{a\eta_q}\; a_2 + M_{a\eta_q}\; C_2 \right ) \non
 && -\frac{3}{2} \xi_t \left [ f_{B_s}F_{a\eta_q} \left( a_7+ a_9 \right )
+ M_{a\eta_q} \; C_{10} + M_{a\eta_q}^{P_2}\; C_{8} \right ], \label{eq:a001}\\
\sqrt{2}{\cal A}({\bar B}_s^0\to\pi^0\eta_s)&=& \xi_u \left ( f_\pi F_{e\eta_s}\; a_2 + M_{e\eta_s}\; C_2\right ) \non
 && -\frac{3}{2} \xi_t \left [ f_\pi F_{e\eta_s} \left (a_9-a_7 \right )
+ M_{e\eta_s} \left ( C_{8} + C_{10} \right ) \right ], \label{eq:a002}
 \eeq
where $\xi_u = V_{ub}V_{us}^*$, $\xi_t = V_{tb}V_{ts}^*$, and $a_i$ are the
combinations of the Wilson coefficients $C_i$ as defined for example in Ref.\cite{fan2013}.

For $\bar {B}_s^0\to \eta \eta, \eta \etar, \etar\etar$ decays, the LO decay amplitudes are
\beq
\sqrt{2}{\cal A}({\bar B}_s^0\to\eta\eta)&=& \cos^2\phi {\cal A}(\eta_q\eta_q)
-\sin(2\phi){\cal A}(\eta_q\eta_s) + \sin^2\phi {\cal A}(\eta_s\eta_s),  \\
{\cal A}({\bar B}_s^0\to\eta\etar)&=& \left [ {\cal A}(\eta_q\eta_q)
-{\cal A}(\eta_s\eta_s)\right ] \cos\phi\sin\phi +\cos(2\phi){\cal A}(\eta_q\eta_s),  \\
\sqrt{2} {\cal A}({\bar B}_s^0\to\etar \etar)&=& \sin^2\phi {\cal A}(\eta_q\eta_q)
+\sin(2\phi){\cal A}(\eta_q\eta_s) + \cos^2\phi {\cal A}(\eta_s\eta_s),
\eeq
with
\beq
    {\cal A}({\bar B}_s^0\to\eta_q\eta_q)&=& \xi_u\; M_{a\eta_q}\; C_2
    -\xi_t \; M_{a\eta_q}\left ( 2C_4 + 2C_6 + \frac{1}{2}C_8 +
\frac{1}{2}C_{10} \right),
\label{eq:a003} \\
\sqrt2{\cal A}({\bar B}_s^0\to\eta_q\eta_s)&=& \xi_u \left ( f_q F_{e\eta_s}\; a_2 + M_{e\eta_s}\; C_2 \right)
-\xi_t \left [ f_q F_{e\eta_s}\left ( 2a_3-2a_5-\frac{1}{2}a_7+\frac{1}{2}a_9 \right) \right. \non&& \left. +
M_{e\eta_s}\left (2C_4+2C_6+\frac{1}{2}C_{8}+\frac{1}{2}C_{10} \right) \right ],
\label{eq:a004} \\
{\cal A}({\bar B}_s^0\to\eta_s\eta_s)&=& -2\xi_t
 \left [ f_s F_{e\eta_s}\left ( a_3+a_4-a_5+\frac{1}{2}a_7-\frac{1}{2}a_9-\frac{1}{2}a_{10}\right) \right. \non &&
\left. + \left ( f_s F_{e\eta_s}^{P_2} +f_{B_s}F_{a\eta_s}^{P_2}\right)\left ( a_6-\frac{1}{2}a_8
\right) \right. \non && \left. + \left ( M_{e\eta_s} +M_{a\eta_s} \right ) \left (C_3+C_4
+C_6-\frac{1}{2}C_8-\frac{1}{2}C_9-\frac{1}{2}C_{10} \right) \right]. \label{eq:a005}
 \eeq
The individual decay amplitudes $(F_{eM_3}, F_{eM_3}^{P2},\cdots)$ in Eqs.~(\ref{eq:a001},\ref{eq:a002},\ref{eq:a003}-\ref{eq:a005}) are obtained by
evaluating the Feynman diagrams in Fig.~1 analytically.
Here $(F_{eM_3}, F_{eM_3}^{P2})$ and $(M_{eM_3},
M_{eM_3}^{P2})$  come from the evaluations of Figs.(1a,1b) and Figs.(1c,1d), respectively; while $(F_{aM_3},
F_{aM_3}^{P2})$ and $(M_{aM_3}, M_{aM_3}^{P2})$ are obtained by evaluating Figs.(1e,1f) and Figs.(1g,1h), respectively.
One can find the expressions of all these decay amplitudes for example in Refs.\cite{pieta,ali07}. For the sake of the reader, we show $F_{eM_3}$ and $F_{eM_3}^{P2}$ explicitly here:
\beq
F_{eM_3}&=& 8\pi C_F M_{B_{s}}^{4} \int_0^1 d x_{1} dx_{3}\, \int_{0}^{\infty} b_1 db_1
 b_3 db_3\, \phi_{B_{s}}(x_1,b_1)\non
&& \cdot \left \{ \left [(1+x_3) \phi_3^A(x_3) + r_3 (1-2x_3)
\left (\phi_3^P (x_3)+\phi_3^T (x_3) \right ) \right ]
\right.\non
&& \left.
\quad \cdot \alpha_s(t_e^1)  h_e(x_1,x_3,b_1,b_3)\exp[-S_{ab}(t_e^1)]
\right.\non
&& \left.
+ 2r_3 \phi_3^P (x_3) \cdot \alpha_s(t_e^2)
 h_e(x_3,x_1,b_3,b_1)\exp[-S_{ab}(t_e^2)] \right\} \;,
\label{eq:ab}
\eeq
\beq
F_{eM_3}^{P_{2}} &=&  16\pi C_F   M_{B_{s}}^{4}
\int_0^1 d x_{1} dx_{3}\, \int_{0}^{\infty} b_1 db_1
 b_3 db_3\, \phi_{B_{s}}(x_1,b_1) r_2\non
&& \cdot  \left \{ \left [\phi_3^A(x_3) + r_3 (2+x_3)\phi_3^P (x_3)- r_3 x_3\phi_3^T (x_3)\right ] \right.\non &&
\left. \quad \cdot  \alpha_s(t_e^1)  h_e(x_1,x_3,b_1,b_3)\exp[-S_{ab}(t_e^1)] \right.\non && \left. + 2 r_3 \phi_3^P
(x_3) \alpha_s(t_e^2)
 h_e(x_3,x_1,b_3,b_1)\exp[-S_{ab}(t_e^2)] \right \} ,
\eeq
where $C_F=4/3$ is the color-factor, $r_2 = m_0^{M_2}/M_{B_{s}} $ and $r_3 = m_0^{M_3}/M_{B_{s}}$ with the
chiral mass $m_0$ for final state meson $M_2$ and $M_3$. The explicit expressions of the hard energy scales
$(t_e^1,t_e^2)$, the hard function $h_e$ and the Sudakov factor $\exp[-S(t)]$ can be found for example in
Refs.~\cite{pieta,ali07}.

\subsection{ NLO contributions}\label{sec:nlo-cont}

\begin{figure}[tb]
\vspace{-5cm}
\centerline{\epsfxsize=18 cm \epsffile{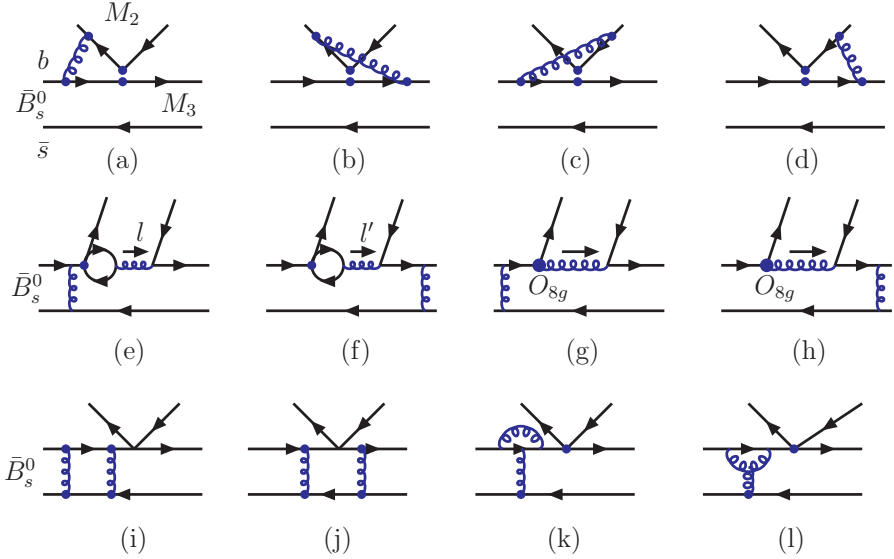}}
\vspace{-13cm}
\caption{Typical Feynman diagrams for NLO contributions:  the vertex corrections (a-d);
the quark-loops (e-f),  the chromo-magnetic penguin contributions (g-h),
and the NLO twist-2 and twist-3 contributions to $B_s \to P$ transition form factors (i-l).}
\label{fig:fig2}
\end{figure}

After many year's efforts, almost all NLO contributions in the pQCD approach
become available now:
\begin{enumerate}
\item[(a)]
The NLO Wilson coefficients $C_i(\mu)$ with $\mu \approx m_b$ \cite{buras96} and the
strong coupling constant $\alpha_s(\mu)$ at two-loop level.

\item[(b)]
The NLO vertex corrections (VC)\cite{npb675}, the NLO contributions from the
quark-loops (QL) \cite{nlo05} or from the chromo-magnetic penguin (MP) operator $O_{8g}$ \cite{o8g2003}.
The relevant Feynman diagrams are shown in Fig.~2(a)-2(h).

\item[(c)]
The NLO twist-2 and twist-3 contributions to the form factors of $B \to P$
transitions (here P refers to the light pseudo-scalar mesons)
\cite{prd85-074004,cheng14a}. Based on the $SU(3)$ flavor symmetry, we will extend directly
the formulaes for NLO contributions to the form factors of $B \to P$ transition
as given in Refs.~\cite{prd85-074004,cheng14a} to the cases for $B_s \to P$ transitions.

\end{enumerate}
In this paper, we adopt the relevant formulaes for all currently known
NLO contributions directly from Refs.~\cite{npb675,nlo05,o8g2003,fan2013,prd85-074004,cheng14a,xiao145}
without further discussion about the details.
The still missing part of the NLO contributions in the pQCD approach is the calculation
for the NLO corrections to the LO hard spectator and annihilation diagrams.
But from the comparative studies for the LO and NLO contributions from different sources in
Refs.~\cite{fan2013,xiao2014}, we believe that those still unknown NLO contributions are most possibly
the higher order corrections to the small LO quantities, and therefore can be neglected safely.

According to Refs.~\cite{npb675,nlo05}, the vertex corrections can be absorbed
into the re-definition of the Wilson coefficients by adding a vertex-function $V_i(M)$ to them.
The expressions of the vertex functions $V_{i}(M)$ can be found easily in Refs.~\cite{npb675,nlo05}.
The NLO "QL" and "MP" contributions  are a kind of penguin correction with the insertion of the
four quark operators and the chromo-magnetic operator $O_{8g}$ respectively,
as shown in Figs.~2(e,f) and 2(g,h).
For the $b\to s$ transition, the relevant effective Hamiltonian $H_{eff}^{ql}$ and $H_{eff}^{mp}$
can be written as the following form:
\beq
H_{eff}^{(ql)}&=&-\sum\limits_{q=u,c,t}\sum\limits_{q{\prime}}\frac{G_F}{\sqrt{2}}
V_{qb}^{*}V_{qs}\frac{\alpha_s(\mu)}{2\pi}C^{q}(\mu,l^2)\left(\bar{b}\gamma_\rho
\left(1-\gamma_5\right)T^as\right)\left(\bar{q}^{\prime}\gamma^\rho
T^a q^{\prime}\right),\label{eq:heff-ql}\\
H_{eff}^{mp} &=&-\frac{G_F}{\sqrt{2}} \frac{g_s}{8\pi^2}m_b\;
V_{tb}^*V_{ts}\; C_{8g}^{eff} \; \bar{s}_i \;\sigma^{\mu\nu}\; (1+\gamma_5)\;
 T^a_{ij}\; G^a_{\mu\nu}\;  b_j, \label{eq:heff-o8g}
\eeq
where $l^2$ is  the invariant mass of the gluon which attaches the quark loops
in Figs.~2(e,f), and the functions $C^{q}(\mu,l^2)$ can be found in Ref.~\cite{nlo05,xiao08b}.
The $C_{8g}^{eff}$ in Eq.~(\ref{eq:heff-o8g}) is the effective Wilson coefficient with
the definition of $C_{8g}^{eff}= C_{8g} + C_5$ \cite{nlo05}.

By analytical evaluations, we find that (a) the decay modes $B_s^0 \to \pi^0 \etap, \eta_q
\eta_q $ and  $\eta_q \eta_s$ do not receive the NLO contributions from the  quark-loop
and the magnetic-penguin diagrams;
and (b) only the $B_s^0 \to \eta_s \eta_s$ decay mode
get the NLO contributions from the quark-loop diagrams and  the $O_{8g}$ operator:
\beq
{\cal M}^{(ql)}_{\eta_s \eta_s}&=& -16m_{B_s}^4\frac{{C_F}^2}{\sqrt{2N_c}} \int_0^1 dx_1dx_2dx_3
\int_0^\infty b_1db_1b_3db_3 \,\phi_{B_s}(x_1)
\Bigl \{ \left [(1+x_3)\phi_{\eta_s}^A(x_2) \phi_{\eta_s}^A(x_3)
\right.
\non && \left.
+ 2 r_{\eta_s}\phi_{\eta_s}^P(x_2) \phi_{\eta_s}^A(x_3)+
r_{\eta_s}(1-2x_3)\phi_{\eta_s}(x_2)(\phi_{\eta_s}^P(x_3)
+ \phi_{\eta_s}^T(x_3)) \right]
\non &&
\cdot \alpha_s^2(t_a) \cdot h_e(x_1,x_3,b_1,b_3)\cdot  \exp\left [-S_{ab}(t_a)\right ]\; C^{(q)}(t_a,l^2)
\non &&
+ 2r_{\eta_s}\phi_{\eta_s}^A(x_2)\phi_{\eta_s}^P(x_3) \cdot
\alpha_s^2(t_b) \cdot h_e(x_3,x_1,b_3,b_1)
\cdot \exp[-S_{ab}(t_b)] \; C^{(q)}(t_b,l'^2)\Bigr \},
\eeq
\beq
{\cal M}^{(mp)}_{\eta_s \eta_s} &=& -32m_{B_s}^6\frac{{C_F}^2}{\sqrt{2N_c}} \int_0^1 dx_1dx_2dx_3
\int_0^\infty b_1db_1b_2db_2b_3db_3\, \phi_{B_s}(x_1)\non
&& \hspace{-1cm}\times \left \{ \left [(1-x_3) \left [ 2\phi_{\eta_s}^A(x_3)+ r_{\eta_s}(3\phi_{\eta_s}^P(x_3)
+\phi_{\eta_s}^T(x_3) )
+ r_{\eta_s}x_3(\phi_{\eta_s}^P(x_3)
-\phi_{\eta_s}^T(x_3))\right] \phi_{\eta_s}^A(x_2) \right.\right. \non
&& \hspace{-1cm} \left.\left.
- r_{\eta_s} x_2(1+x_3) (3\phi_{\eta_s}^P(x_2)
-\phi_{\eta_s}^T(x_2))\phi_{\eta_s}^A(x_3)\right]
\cdot \alpha_s^2(t_a) h_g(x_i,b_i)\cdot \exp[-S_{cd}(t_a)]\; C_{8g}^{eff}(t_a)
\right. \non
&&  \hspace{-1cm} \left.
+ 4r_{\eta_s}\phi_{\eta_s}^A(x_2)\phi_{\eta_s}^P(x_3)
\cdot \alpha_s^2(t_b) \cdot h'_g(x_i,b_i)
\cdot \exp[-S_{cd}(t_b)]\; C_{8g}^{eff}(t_b)\right\},
\eeq
where the terms proportional to small quantity $r^2_{\eta_s}$ are not shown explicitly.
The expressions for the hard functions $(h_e,h_g)$, the functions
$C^{(q)}(t_a,l^2)$ and $C^{(q)}(t_b,l'^2)$, the Sudakov functions $S_{ab,cd}(t)$, the
hard scales $t_{a,b}$ and the effective Wilson coefficients $C_{8g}^{eff}(t)$,
can be found easily for example in Refs.~\cite{nlo05,fan2013,xiao145}.

The NLO twist-2 and twist-3 contributions to the form factors of $B \to \pi$ transition
have been calculated very recently in Refs.~\cite{prd85-074004,cheng14a}.
Based on the $SU(3)$ flavor symmetry, we extend the formulaes of NLO contributions for $B\to \pi$
transitions form factor as given in Refs.~\cite{prd85-074004,cheng14a} to the cases for
$B_s \to (\pi,\eta_q,\eta_s)$ transition form factors directly, after making appropriate
replacements for some parameters.
The NLO form factor $f^+(q^2)$ for $B_s \to \pi$ transition, for example, can be
written as the form of
\beq
f^+(q^2)|_{\rm NLO} &=& 8 \pi m^2_{B_s} C_F \int{dx_1 dx_2} \int{b_1 db_1 b_2 db_2}
\phi_{B_s}(x_1,b_1)\non &&
\hspace{-2cm}\times \Biggl \{ r_\pi \left [\phi_{\pi}^{P}(x_2) - \phi_{\pi}^{T}(x_2) \right ]
\cdot \alpha_s(t_1)\cdot e^{-S_{B_s\pi}(t_1)}\cdot S_t(x_2)\cdot h(x_1,x_2,b_1,b_2) \non
&&\hspace{-2cm}  + \Bigl [ (1 + x_2 \eta)
\left (1 + F^{(1)}_{\rm T2}(x_i,\mu,\mu_f,q^2)\; \right ) \phi_{\pi}^A(x_2)
+ 2 r_\pi \left (\frac{1}{\eta} - x_2 \right )\phi_{\pi}^T(x_2)
- 2x_2 r_\pi \phi_{\pi}^P(x_2) \Bigr ] \non
&& \hspace{-1cm} \cdot \alpha_s(t_1)\cdot e^{-S_{B_s\pi}(t_1)} \cdot S_t(x_2)\cdot h(x_1,x_2,b_1,b_2)\non
&& \hspace{-2cm} + 2 r_{\pi} \phi_{\pi}^P(x_2) \left (1 + F^{(1)}_{\rm T3}(x_i,\mu,\mu_f,q^2)\right )
\cdot \alpha_s(t_2)\cdot e^{-S_{B_s\pi}(t_2)} \cdot S_t(x_2)\cdot h(x_2,x_1,b_2,b_1) \Biggr \},
\label{eq:ffnlop}
\eeq
where $\eta=1-q^2/m_{B_s}^2$ with $q^2=(P_{B_s}-P_3)^2$ and $P_3$ is the momentum of the 
meson $M_3$ which absorbed the spectator light quark of the B meson, $\mu$ ($\mu_f$) is  the
renormalization (factorization ) scale, the hard scale $t_{1,2}$ are
chosen as the largest scale of the propagators in the hard $b$-quark decay diagrams
\cite{prd85-074004,cheng14a}, the
function $S_t(x_2)$ and the hard function $h(x_i,b_j)$ can be found in
Refs.~\cite{prd85-074004,cheng14a}.
And finally the NLO factor $F^{(1)}_{\rm T2}(x_i,\mu,\mu_f,q^2)$ and
$F^{(1)}_{\rm T3}(x_i,\mu,\mu_f,q^2)$ which describe the NLO
twist-2 and twist-3 contribution to the form factor $f^{+,0}(q^2)$ of the 
$B_s \to \pi$ transition can be found in Refs.~\cite{prd85-074004,cheng14a,xiao145}.
For $B_s \to \pi$ transition, for example, these two factors can be written as:
\beq
F^{(1)}_{\rm T2}&=& \frac{\alpha_s(\mu_f) C_F}{4 \pi}
\Biggl [\frac{21}{4} \ln{\frac{\mu^2}{m^2_{B_s}}}
-(\frac{13}{2} + \ln{r_1}) \ln{\frac{\mu^2_f}{m^2_{B_s}}}
+\frac{7}{16} \ln^2{(x_1 x_2)}+ \frac{1}{8} \ln^2{x_1} \non
&&+ \frac{1}{4} \ln{x_1} \ln{x_2}
+ \left (- \frac{1}{4}+ 2 \ln{r_1} + \frac{7}{8} \ln{\eta} \right ) \ln{x_1}
+ \left (- \frac{3}{2} + \frac{7}{8} \ln{\eta} \right) \ln{x_2} \non
&&+ \frac{15}{4} \ln{\eta} - \frac{7}{16} \ln^{2}{\eta}
+ \frac{3}{2} \ln^2{r_1} - \ln{r_1}
+ \frac{101 \pi^2}{48} + \frac{219}{16} \Biggr ],  \label{eq:ffnlot2}\\
F^{(1)}_{\rm T3}&=&\frac{\alpha_s(\mu_f) C_F}{4 \pi}
\Biggl [\frac{21}{4} \ln{\frac{\mu^2}{m^2_{B_s}}}
- \frac{1}{2}(6 + \ln{r_1}) \ln{\frac{\mu^2_f}{m^2_{B_s}}}
+ \frac{7}{16} \ln^2{x_1} - \frac{3}{8} \ln^2{x_2} \non
&& \hspace{-1cm}+ \frac{9}{8} \ln{x_1} \ln{x_2}
+ \left (- \frac{29}{8}+ \ln{r_1} + \frac{15}{8} \ln{\eta} \right ) \ln{x_1}
+ \left (- \frac{25}{16} + \ln{r_2} + \frac{9}{8} \ln{\eta} \right) \ln{x_2} \non
&&\hspace{-1cm}+ \frac{1}{2} \ln{r_1} - \frac{1}{4} \ln^{2}{r_1} + \ln{r_2}
- \frac{9}{8} \ln{\eta} - \frac{1}{8} \ln^{2}{\eta} + \frac{37 \pi^2}{32}
+ \frac{91}{32} \Biggr ],
\eeq
where $r_i=m^2_{B_s}/\xi_i^2$ with the choice of $\xi_1=25 m_{B_s}$
and $\xi_2=m_{B_s}$. For the considered $B_s \to (\pi^0,\eta^{(')} )\; \eta^{(')}$ decays,
the large recoil region corresponds to the energy fraction  $\eta  \sim \textit{O}(1)$.
We also set $\mu=\mu_f=t$ in order to minimize the NLO contribution to the form factors
\cite{prd83-054029,cheng14a}.

%%---------------------------------------------------------------

\section{Numerical results}\label{sec:n-d}

In the numerical calculations the following input parameters (here the masses, decay constants and QCD scales are in
unit of GeV) will be used \cite{hfag2012,pdg2012}:
 \beq
  \Lambda_{\overline{\mathrm{MS}}}^{(5)} &=& 0.225, \quad
f_{B_{s}} =0.23, \quad f_{\pi} = 0.13,\quad m_{B_s} =  5.37,\quad  m_\eta=0.548, \quad m_{\etar}=0.958, \non \quad
m_0^\pi &=& 1.4, \quad \tau_{B_s^0} = 1.497 \ \ {\rm ps}, \quad m_b=4.8 , \quad M_W = 80.41. \label{eq:para}
 \eeq
For the CKM matrix elements, we adopt the Wofenstein parametrization and
use the following CKM parameters: $\lambda = 0.2246$, $ A = 0.832$, $\bar{\rho} = 0.130 \pm 0.018$
and $\bar{\eta} = 0.350 \pm 0.013$.

%%\subsection{Form factors at LO and NLO level}

Taking $B_s \to \pi$  transition as an example, we calculate and present the pQCD predictions
for the form factors $F_0^{\bar B^0_s \to \pi}(0)$ at the LO and NLO level respectively:
\beq
F_0^{\bar B^0_s \to \pi}(0) =  \left\{ \begin{array}{cc}
0.22\pm 0.05, & {\rm LO}, \\ 0.24\pm 0.05, & {\rm NLO}, \\ \end{array} \right.
\eeq
where the error comes from the uncertainty of $\omega_{B_s}=0.50\pm 0.05$ GeV,
$f_{B_s}=0.23\pm 0.02$ GeV and the Gegenbauer moments $a_2^\pi = 0.44 \pm 0.22$.
Explicit calculations tell us that the NLO twist-2 enhancement to the full LO prediction
is around $25\%$, but it is largely canceled
by the negative NLO twist-3 contribution and finally lead to a small total  enhancement
(about $7\% \sim 9\%$) to the full LO prediction, as predicted in Ref.~\cite{cheng14a}.

%%\subsection{Branching Ratios}

For the considered five $\bar{B}_s^0$ decays,  the CP-averaged branching ratios
can be written in the following form:
\beq
{\rm Br}(B_s^0\to f) = \frac{G_F^2 \tau_{B_s}}{32\pi m_{B_s}} \;
 \frac{1}{2}  \Bigl[  |{\cal A}(\bar{B}_s^0\to f)|^2
+|{\cal A}(B_s^0\to \bar{f})|^2 \Bigr ],
 \label{eq:br0}
\eeq
where $\tau_{B_s}$ is the lifetime of the $B_s^0$ meson.

In Table I, we list the pQCD predictions for the CP-averaged branching ratios
of the considered $B_s^0$ decays.
The label ``NLO-I" means that all currently known NLO contributions are taken into account except for those to the
form factors. As a comparison, we also show the central values of the LO pQCD predictions as given in
Ref.~\cite{ali07}, the partial NLO predictions in Ref.~\cite{xiao08a}
and the QCDF predictions in Ref.~\cite{npb675} in last three columns of Table \ref{tab:br1}.
The main theoretical errors come from the uncertainties of the various input parameters: such
as $\omega_{B_s}=0.50 \pm 0.05$, $f_{B_s}=0.23\pm 0.02$ GeV and $a_2^\pi=0.44\pm 0.22$.
The total errors of our pQCD¡¡predictions are obtained by adding the individual errors in quadrature.

\begin{table}
\caption{ The pQCD predictions for the branching ratios ( in unit of $10^{-6}$ ) of the considered five
$\bar{B}_s^0$ decays. As a comparison, we also list the theoretical
predictions as given in Refs.~\cite{ali07,xiao08a,npb675}, respectively.}
\label{tab:br1}
%%\begin{tabular*}{l|c| lll|cc} \hline \hline
\begin{tabular*}{12cm}{@{\extracolsep{\fill}}l|lll|ccc} \hline\hline
 Mode & LO \qquad \qquad& NLO-I \qquad& NLO \qquad& LO \cite{ali07}& NLO-I\cite{xiao08a}& QCDF \cite{npb675}
 \\ \hline
 $\bar B_s^0\to \pi^0\eta $&    $0.05  $&$0.05  $&$0.06\pm 0.03        $&$0.05 $&$0.03  $&$0.08 $\\
 $\bar B_s^0\to \pi^0\etar$&    $0.10  $&$0.11  $&$0.13\pm 0.06        $&$0.11 $&$0.08  $&$0.11 $\\
 $\bar B_s^0\to \eta\eta  $&    $10.1  $&$9.9   $&$10.6^{+3.8}_{-2.7}  $&$8.0  $&$10.0  $&$15.6  $\\
 $\bar B_s^0\to \eta\etar $&    $27.5  $&$38.4  $&$41.4^{+16.4}_{-12.0}$&$21.0 $&$34.9  $&$54.0 $\\
 $\bar B_s^0\to \eta'\etar$&    $20.5  $&$37.7  $&$41.0^{+17.5}_{-13.4}$&$14.0 $&$25.2  $&$41.7 $\\
\hline\hline

%\end{tabular}
\end{tabular*}
\end{table}

From the numerical results as listed in Table \ref{tab:br1}, one can observe the following points
\begin{itemize}
\item
For $\bar B_s^0\to (\pi^0\eta,\pi^0\etar,\eta\eta)$ decays, the NLO enhancements to the full LO
predictions are small in size: less than $30\%$.
For $\bar B_s^0\to (\eta\etar,\etar\etar)$ decays, however, the NLO enhancements can be
as large as $100\%$. The branching ratios at the order of $4\times 10^{-5}$ should be measured
at LHCb or super-B factory experiments.

\item
By comparing the numerical results as listed in the third (NLO-I) and fourth (NLO) column,
one can see that the  NLO contributions to the form factors along
can provide $\sim 10\%$ enhancement to the branching ratios.

\item
The pQCD predictions agree with the QCDF predictions within one standard deviation.
The pQCD predictions given in some previous works \cite{ali07,xiao08a} are confirmed
by our new calculations. Some differences between the central values
are induced by the different choices of some input parameters, such as the Gagenbauer moments and
the CKM matrix elements.

\item
The main theoretical errors are coming from the uncertainties of input parameters
$\omega_{B_s}=0.50\pm 0.05$, $f_{B_s}=0.23\pm 0.02$ GeV  and $a_2^\pi=0.44\pm 0.22$.
The total theoretical error is in general around $30\%$ to $50\%$.

\end{itemize}

Now we turn to the evaluations of the CP-violating asymmetries of the five considered decay modes.
In the  $B_s$ system, we expect a much larger decay width difference: $\Delta\Gamma_s/(2\Gamma_s)\sim
-10\% $ \cite{hfag2012}. Besides the direct CP violation $\cala_f^{dir}$, the CP-violating asymmetry
$S_f$ and $H_f$ are defined as usual \cite{ali07,xiao08a}
 \beq
\cala_f^{dir}=\frac{|\lambda|^2-1 }{1+|\lambda|^2},\quad
{\cal S}_f=\frac{2 {\rm Im}[\lambda]}{1+|\lambda|^2},\quad
\calh_f=\frac{2 {\rm Re}[\lambda]}{1+|\lambda|^2}.
\eeq
They satisfy the normalization relation $|\cala_f|^2+|{\cal S}_f|^2+|\calh_f|^2=1$, while 
the parameter $\lambda$ is of the form 
\beq
\lambda=\eta_f e^{2i\epsilon}\frac{A(\ov B^0_s \to f)}{A(B^0_s \to \bar f)},
\eeq
where $\eta_f$ is $+1(-1)$ for a CP-even(CP-odd) final state f and $\epsilon
=\arg[-V_{ts}V_{tb}^*]$ is very small in size.

The pQCD predictions for the direct CP asymmetries $\cala_f^{dir}$, the mixing-induced CP
asymmetries $S_f$ and $H_f$ of the considered decay modes are listed in Table
\ref{tab:acp1} and Table \ref{tab:acp2}. In these tables, the label ``LO" means the
LO pQCD predictions, the label ``$+$VC", ``$+$QL", ``$+$MP", as well as ``NLO" means
that the contributions from the vertex corrections, the quark loops, the magnetic penguins, 
and all known NLO contributions are added to the LO results, respectively.
As a comparison, the LO pQCD predictions as given in Ref.~\cite{ali07}
and the QCDF predictions in Ref.~\cite{npb675} are also listed in Table \ref{tab:acp1}
and \ref{tab:acp2}. The errors here are defined in the same way as for the branching ratios.

\begin{table}
\caption{ The pQCD predictions for the direct CP asymmetries (in $\%$) of the five $\bar B_s^0$ 
decays. The meaning of the labels are described in the text. }
 \label{tab:acp1}
\begin{tabular*}{15cm}{@{\extracolsep{\fill}}l|llccc|cc} \hline\hline
 Mode &{\rm LO}&{\rm + VC}&{\rm +QL}&{\rm +MP}
 &{\rm NLO}&{\rm pQCD}\cite{ali07}&{\rm QCDF}\cite{npb675} \\ \hline
 $\bar B_s^0\to \pi^0\eta $&$-2.5^{+8.9}_{-7.8} $&$39.8$ &$-    $&$-    $&$40.3^{+5.4}_{-7.5}$&$-0.4^{+0.3}_{-0.3}$&$-$\\
 $\bar B_s^0\to \pi^0\etar$&$24.7^{+0.3}_{-1.0} $&$52.7$ &$-    $&$-    $&$51.9^{+2.9}_{-3.3}$&$20.6^{+3.4}_{-2.9}$&$27.8^{+27.2}_{-28.8}$\\
 $\bar B_s^0\to \eta\eta $ &$-0.2^{+0.3}_{-0.2} $&$-2.2$ &$1.7 $&$-1.8$&$-2.3^{+0.5}_{-0.4}$&$-0.6^{+0.6}_{-0.5}$&$-1.6^{+2.4}_{-2.4}$\\
 $\bar B_s^0\to \eta\etar $&$-1.1\pm 0.1 $       &$-1.0$ &$0.1 $&$-0.1$&$-0.2\pm 0.2$&$-1.3^{+0.1}_{-0.2}$&$0.4^{+0.5}_{-0.4}$\\
 $\bar B_s^0\to \eta'\etar$&$1.4\pm 0.2  $       &$ 1.5$ &$2.7  $&$2.8 $&$2.8\pm 0.4 $&$1.9^{+0.4}_{-0.5} $&$2.1^{+1.3}_{-1.4}$\\
\hline\hline
\end{tabular*} \end{table}

\begin{table}
\caption{The pQCD predictions for the mixing-induced CP asymmetries (in $\%$) $S_f$(the first row) 
and $H_f$ (the second row). The meaning of the labels are the same as in Table~\ref{tab:acp1}. } 
\label{tab:acp2}
\begin{tabular*}{15cm}{@{\extracolsep{\fill}}l|ccccc|c} \hline\hline
 Mode & LO &  + VC& +QL & +MP  &  NLO& pQCD \cite{ali07}
 \\ \hline
 $\bar B_s^0\to \pi^0\eta $&$13.7^{+6.6}_{-8.3} $ &$11.3$&$-     $&$-     $&$8.0^{+1.8}_{-2.7}  $&$17^{+18}_{-13}$\\
                           &$99.0^{+0.5}_{-1.4} $ &$91.0 $&$-     $&$-    $&$91.2^{+3.0}_{-2.4} $&$99\pm 1$\\
$\bar B_s^0\to\pi^0\etar  $&$-22.2^{+10.0}_{-7.3}$&$-24.9$&$-    $&$-    $ &$-24.9^{+9.5}_{-6.1}$&$-17^{+8}_{-9}$\\
                           &$94.3^{+2.0}_{-1.8} $&$81.3 $&$-     $&$-    $ &$81.8^{+0.5}_{-0.1} $&$96^{+2}_{-2}$\\
$\bar B_s^0\to\eta\eta    $&$-0.6^{+0.4}_{-0.3} $&$2.7  $&$-1.8 $&$-2.2 $&$-2.2^{+0.6}_{-0.5} $&$3.0^{+1}_{-1}$\\
                           &$100.0 $             &$99.9$&$100.0$&$100.0 $&$99.9$&$100.0$\\
$\bar B_s^0\to\eta\etar   $&$-0.1\pm 0.1 $       &$-0.4 $&$0.1  $&$0.1  $&$0.1\pm 0.2 $&$4.0$\\
                           &$100.0$              &$100.0$&$100.0$&$100.0$&$100.0$&$100.0$\\
$\bar B_s^0\to\etar\etar  $&$0.8\pm 0.1  $       &$1.8  $&$2.0  $&$2.5  $&$2.5^{+0.2}_{-0.4}  $&$4.0^{+1.0}_{-1.0}$\\
                           &$100.0$              &$100.0$&$99.9$&$99.9$  &$99.9$&$100.0$\\
 \hline\hline
\end{tabular*} \end{table}

From the pQCD predictions for the CP violating asymmetries of the five
considered $\bar{B}_s$ decays as listed  in the Table~\ref{tab:acp1} and \ref{tab:acp2}, one can see the following points:
\begin{itemize}
\item
For $\bar{B}^0_s \to (\eta\eta,\eta\etar,\etar\etar)$ decays, the pQCD predictions for $\cala_f^{dir}$ and
${\cal S}_f$ are very small: less than $3\%$  in magnitude. The NLO effects are in fact also negligibly
small.

\item
For $\bar{B}^0_s \to (\pi^0\eta,\pi^0\etar)$ decays, however, the NLO
pQCD predictions for $\cala_f^{dir}$ can be as large as $40\% - 52\%$.
The NLO contributions can provide large enhancements to the LO pQCD predictions for   $\cala_f^{dir}$.
Since the branching ratios of $\bar{B}^0_s \to (\pi^0\eta,\pi^0\etar)$ decays are at the $10^{-8}$ level,
unfortunately, there is no hope to observe their CP violation even at Super-B factory experiments.

\end{itemize}

\section{SUMMARY}

In short, we calculated the  branching ratios and CP-violating asymmetries of
the five $\bar B_s^0\to (\pi^0,\etap) \etap$ decays by employing the pQCD factorization approach.
All currently known NLO contributions, specifically those NLO twist-2 and twist-3
contributions to the relevant form factors, are taken into account.
From our studies, we found the following results:
\begin{itemize}
\item
For $\bar{B}_s^0\to (\eta\etar,\etar\etar)$ decays, the NLO enhancements to their branching ratios 
can be as large as $100\%$. For other three decay modes, however, the NLO enhancements  
are less than $30\%$.
The  newly known NLO twist-2 and twist-3 contributions to the form factors along
can provide $\sim 10\%$ enhancements to the branching ratios.

\item
For the $\bar{B}_s \to \pi^0 \etap$ decays, the LO pQCD predictions for 
$\cala_f^{dir}$ can be enhanced significantly by the inclusion of the
NLO contributions. For other three decays, the NLO contributions are small in size.

\item
For $\bar{B}_s \to (\eta \etap,\etar\etar)$ decays, their branching ratios are at the order
of $4\times 10^{-5}$, which may be measurable at LHCb or super-B factory experiments.

\end{itemize}

\begin{acknowledgments}
The authors would like to thank Cai-Dian L\"u and Xin Liu for helpful discussions.
This work was supported by the
National Natural Science Foundation of China under Grant No.~11235005.

\end{acknowledgments}

%%%%%%%%%%%%%%%%%%%%%%%%%%%%%%%%%%%%%%%%%%%%%%%%%%%%%%%%%%%%%%%%%%%%%%%%%%%%%%%%%%%%%%%%%%%%%%5
%                                 reference
%%%%%%%%%%%%%%%%%%%%%%%%%%%%%%%%%%%%%%%%%%%%%%%%%%%%%%%%%%%%%%%%%%%%%%%%%%%%%%%%%%%%%%%%%%%%%%%%%

%\newpage

\end{document}